\documentclass[prl,superscriptaddress,twocolumn]{revtex4-1}
\usepackage{graphicx}
\usepackage{amsmath}
\usepackage{amssymb}
\usepackage{gensymb}
\usepackage[margin=2cm]{geometry}
\usepackage{setspace}

\newcommand{\reftextit}[1]{}

\newcommand{\ket}[1]{\left| #1 \right>}

\newcommand{\braket}[2]{\left<#1\vphantom{#2}\hspace{-1mm}\right.\left|#2\vphantom{#1}\right>}
\begin{document}
\title{Localized inter-valley defect excitons as single-photon emitters in WSe$_2$}

\author{Lukas Linhart}
\affiliation{Institute for Theoretical Physics, Vienna University of Technology, 1040 Vienna, Austria, EU.}
\author{Matthias Paur}
\affiliation{Institute of Photonics, Vienna University of Technology, 1040 Vienna, Austria, EU.}
\author{Valerie Smejkal}
\affiliation{Institute for Theoretical Physics, Vienna University of Technology, 1040 Vienna, Austria, EU.}
\author{Joachim Burgd\"orfer}
\affiliation{Institute for Theoretical Physics, Vienna University of Technology, 1040 Vienna, Austria, EU.}
\author{Thomas Mueller}
\affiliation{Institute of Photonics, Vienna University of Technology, 1040 Vienna, Austria, EU.}
\author{Florian Libisch}
\affiliation{Institute for Theoretical Physics, Vienna University of Technology, 1040 Vienna, Austria, EU.}

\begin{abstract}
  Single-photon emitters play a key role in present and emerging quantum
  technologies. Several recent measurements have established monolayer
  WSe$_2$ as a promising candidate for a reliable single photon
  source. The origin and underlying microscopic processes have
  remained, however, largely elusive. We present a multi-scale
  tight-binding simulation for the optical spectra of WSe$_2$ under
  non-uniform strain and in the presence of point defects employing
  the Bethe-Salpeter equation. Strain locally shifts excitonic energy
  levels into the band gap where they overlap with localized intra-gap
  defect states. The resulting hybridization allows for efficient
  filing and subsequent radiative decay of the defect states. We
  identify inter-valley defect excitonic states as the likely
  candidate for anti-bunched single-photon emission. This proposed
  scenario is shown to account for a large variety of experimental
  observations including brightness, radiative transition rates, the variation of the
  excitonic energy with applied magnetic and electric fields as well
  as the variation of the polarization of the emitted photon with the
  magnetic field.
\end{abstract}
\maketitle

\hyphenation{hetero-bi-layers}

Transition Metal Dichalcogenides (TMDs) have attracted considerable
interest over the last decade. A direct band gap in the mono layer
case \cite{Mak2010,Splendiani2010}, extremely large excitonic
binding energies in the order of 300-500~meV \cite{He2014,Chernikov2014,Cao2012} and
valley as well as spin selective optical transitions due to the
$\rm{D}_{3\rm{h}}$ symmetry make these materials very promising
candidates for optical devices \cite{Mueller2018,Wang2017}.  Single
photon emitters (SPEs) in $\rm{WSe}_2$ are among the most intriguing
candidates for such future optical applications attracting
considerable attention in the field of two-dimensional materials
\cite{Tonndorf2015,Kumar2015,He2015,Srivastava2015,Kumar2016,Schwarz2016,
Berraquero2016,Kern2016,Branny2017,Luo2018,Chakraborty2019,
Lu2019,Koperski2015,Chakraborty2015,Berraquero2017,Clark2016,He2016,Chakraborty2017}.
Single-photon emitters promising photon emission ``on demand'' are key
building blocks for optoelectronic and photonic-based
quantum-technological devices, e.g., for generating entangled
photons \cite{Huber2017}.

SPEs in WSe$_2$ emit antibunched light from highly localized spots in
suspended WSe$_{2}$ flakes featuring a narrow linewidth (down to 100
$\mu eV$) and an intricate fine structure (for a review
see~\cite{Koperski2017}). A large number of experimental
investigations have provided key insight to help unraveling the puzzle
of the microscopic origin of SPEs. The prominent observation of SPEs
in regions of enhanced strain, for example close to pillars suspending
the WSe$_2$ membrane
\cite{Kern2016,Branny2017,Berraquero2017,Luo2018}, points to the
crucial role of locally non-uniform strain. The large defect density
in WSe$_2$ also seems to play a role in the formation of SPEs
\cite{Luo2018}.  The appearance of doublets in the optical spectra --
i.e., single photon emission lines with energy spacing up to 1 meV --
has been attributed to the exchange interaction between excitons but
the underlying mechanism has remained an open question. While in some early
studies few SPEs were found to be only weakly dependent on the magnetic
field, in most measurements an unexpectedly large effective g-factor
ranging from 8 to 13 was observed
\cite{Kumar2015,He2015,Srivastava2015,Lu2019,Koperski2015,Chakraborty2015,Chakraborty2017}.
Several groups observed bi-exciton doublets with a zero field
splitting in the range of 0.2 to 1 meV
\cite{Kumar2015,He2015,Srivastava2015,Kumar2016,Lu2019,Schwarz2016,
Branny2017,Chakraborty2015,Berraquero2017,Clark2016,He2016}.
For SPEs emitting from the same region, measurements find correlated
polarizations, some preferentially parallel to each other
\cite{Kumar2016,He2015,He2016}, while others feature pairs with orthogonal
polarization, in particular for doublets
\cite{He2015,Lu2019,Schwarz2016,Clark2016,He2016}. Equally
puzzling, both linear and quadratic Stark shifts with applied electric
field were recently found for different SPEs
\cite{Chakraborty2017,Schwarz2016,Chakraborty2019}. On an even more fundamental
level, there is no clear picture as to why a SPE in such a
nanostructure possesses a brightness large enough to be measured at
all. The latter suggests a remarkably large optical transition rate of
the emitting state and a highly efficient repopulation subsequent to
the photon emission.

\begin{figure}
\centering \includegraphics[width=0.43\textwidth]{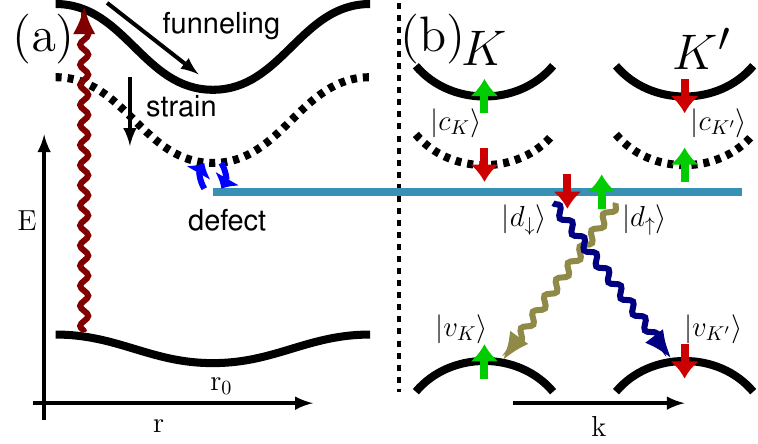}
\caption{Single photon emitter in WSe$_2$, schematically (a) Real
  space representation: a free exciton is created (dark red arrow),
  strain efficiently funnels excitons with the electron in the bright
  (solid black line) and dark (dashed line) conduction band state down
  in energy towards the strain maximum near $r_0$ due to the
  strain-dependent bandgap: mixing of the strain-localized dark
  exciton with a defect state leads to the formation of a strongly
  localized defect exciton.  (b) Reciprocal space: while
  strain-localized exciton states (dashed) remain dark, a point defect
  (horizontal cyan line) breaks the valley selectivity and leads to
  efficient photoemission (dark blue/yellow arrows).}
\label{fig:fig1}
\end{figure}

A detailed microscopic model of the processes involved has remained
elusive. In this letter we present a multi-scale simulation for
WSe$_2$ with locally varying strain and in the presence of point
defects. We employ a tight-binding model for the electronic structure
on a single-particle level and a Bethe-Salpeter approach to account
for two-particle interaction effects. From this simulation the
following microscopic scenario for the origin of SPEs emerges:
strongly non-uniform strain variations (e.g. near the tip of pillars
\cite{Branny2017}) result in the lowering of excitonic energies in the
strained region \cite{Desai14} forming a weakly localized exciton
[Fig.~\ref{fig:fig1}(a)]. In the presence of a point defect in this
region, hybridization with a strongly localized defect level in the
band gap leads to the formation of a novel electron-hole pair
configuration termed inter-valley defect exciton for which the broken
valley symmetry allows efficient radiative decay
[Fig.~\ref{fig:fig1}(b)], the key prerequisite for a SPE. Within this
scenario we are able to quantitatively reproduce measurements of the
SPE fine structure, magnetic- and electric-field behavior as well as
the polarization of the emitted light. The present simulation provides
the theoretical underpinning of previously suggested qualitative
models \cite{Branny2017} and a consistent guide through a diverse
array of seemingly contradictory observations.

\begin{figure}
    \centering
    \includegraphics[width=0.49\textwidth]{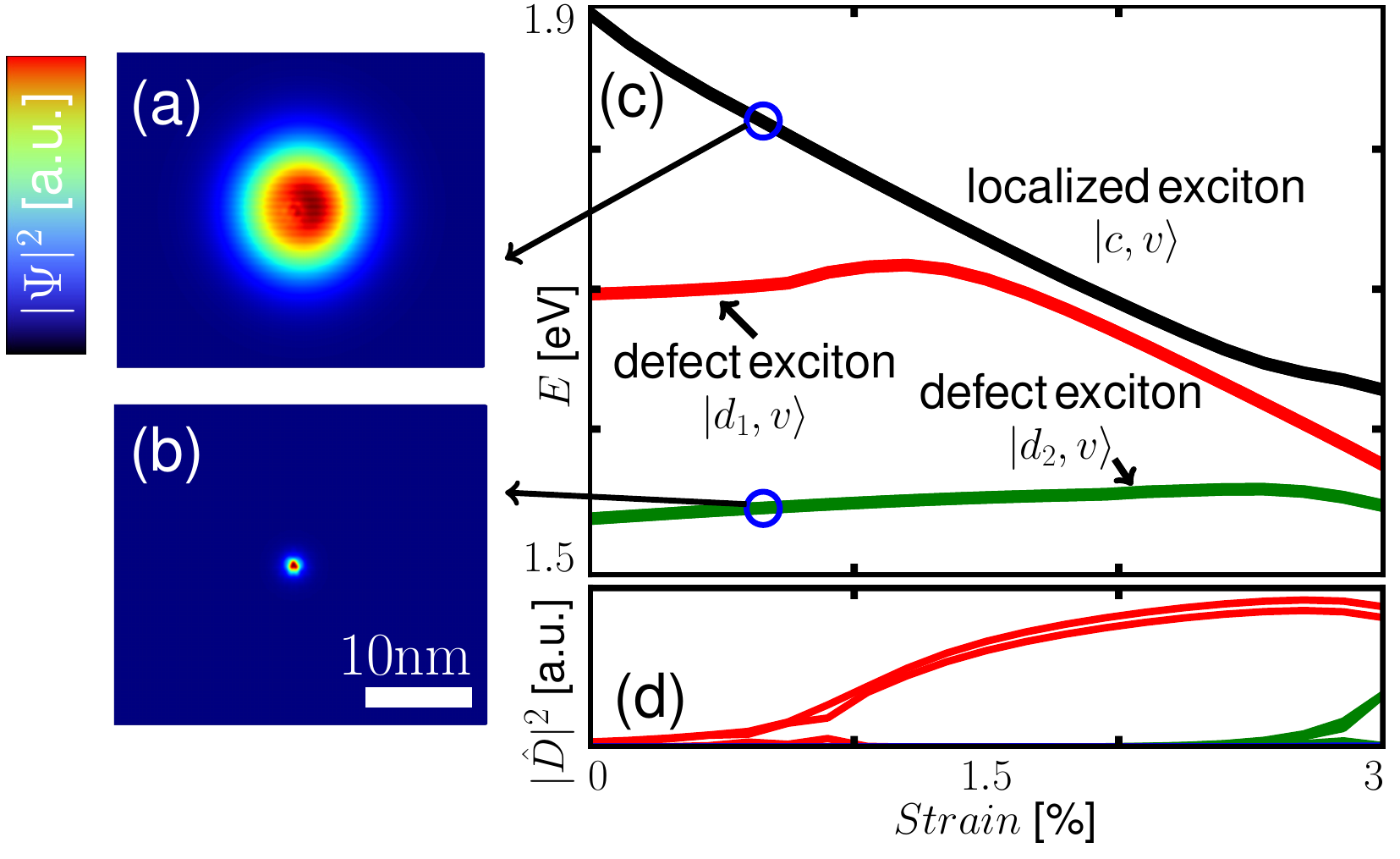}
    \caption{ Wave function and energies of strain-localized states
      within a single-particle picture: (a) and (b): Real space
      representation of a localized conduction band state $\ket c$ and
      strongly localized vacancy defect state $\ket d$. (c) energy
      shift and mixing between localized exciton $\ket{c,v}$ and
      defect excitons $\ket{d_1,v}$ (red) and $\ket{d_2,v}$ (green) as
      a function of strain. (d) Optical oscillation strength of
      excitonic states as function of strain. When $\ket{c,v}$ and
      $\ket{d_i,v}$, $i=1,2$, approach each other the $\ket{d_i,v}$
      exciton becomes bright. Each defect exciton spans a 4$\times$4
      subspace [Eq.~(\ref{space_b})] with two optically active
      transitions (double lines).}
    \label{fig:fig2}
\end{figure}

The starting point of our description on the single-particle level is
a multi-scale approach employing density functional theory (DFT)
calculations \cite{Kresse1993,Kresse1994} to determine the input
parameters of a subsequent tight-binding simulation of large (30 000
atoms) non-uniformly strained WSe$_2$ crystals. This approach
circumvents the need for fitting parameters by projecting onto Wannier
orbitals \cite{Marzari1997,Souza2001} at different strain amplitudes,
and interpolating the tight-binding interactions for locally varying
strain configurations [for details see the supplementary material
  (SM)]. The resulting single-particle eigenstates feature, indeed,
the lowest conduction band states $\ket{c}$ and, consequently, also
the corresponding excitonic states to be localized near the local
maximum of the strain amplitude [Fig.~\ref{fig:fig2}(a) and SM]. The
larger the strain, the more deeply the states get trapped near the
center of the strain pattern [Fig.~\ref{fig:fig2}(c)].  We find
$s$-like radially symmetric conduction ($\ket{c}$) and valence
($\ket{v}$) states that are two fold (valley) degenerate
[Fig.~\ref{fig:fig2}(a)]. The spatial variation of the energy of the
conduction band and excitonic states [Fig.~\ref{fig:fig1}(a)] due to
strain suggests the ``funneling'' of conduction band occupation into
these strain-localized excitonic states \cite{Branny2017}. The present
results are found to be largely independent of the details of the
strain pattern as long as the local variation of strain is
sufficiently smooth such that inter-valley scattering remains
negligible. The highest-lying valence states are spin-polarized
$\braket{v_K}{\uparrow}\approx 1$ and largely consist of atomic
tungsten ${d}_{{x}^2-{y}^2}$ and ${d}_{xy}$ orbitals while the
lowest-lying spin-polarized conduction band states
$\braket{c_K}{\downarrow}\approx 1$ are spanned by tungsten $d_{z^2}$
orbitals \cite{Cappelluti2013}.  These states can be clearly
associated with a well-defined valley quantum number, showing that the
valley symmetry is preserved for these states under strain. Our model
thus reproduces the well-known spin-valley locking in TMDs
\cite{Cao2012}.
Spin-valley locking strongly influences the optical properties of
WSe$_2$: in contrast to molybdenum based TMDs, the exciton in WSe$_2$ 
is ``dark'' since optical intra-valley transitions are spin forbidden, 
with the spin allowed transition several 
tens of meV higher in energy, while inter-valley 
transitions (e.g., $K \rightarrow K'$) are valley 
forbidden \cite{Echeverry2016} [Fig.~\ref{fig:fig1}(b)]. Consequently, direct
optical transition from the excitonic to the ground state are
blocked, raising intriguing questions as to the origin of the observed
strong single-photon emission.

Unraveling the spin-valley locking by a local symmetry breaking
through the ubiquitous presence of defects appears key to
understanding and describing SPEs in WSe$_2$. We therefore include in
our simulations simple prototypical point defects, specifically either
a single or a double Se vacancy. While both break the in-plane
translational symmetry, single Se vacancies also break the out-of
plane inversion symmetry while the latter is preserved by the double
vacancy.  The simulation yields a strongly localized defect state
$\ket{d}$ [Fig.~\ref{fig:fig2}(b)], with energy below the conduction
band, i.e., an electron-like state at normal doping
[Fig.~\ref{fig:fig2}(c)]. It features two spin states
($\ket{d_\uparrow}, \ket{d_\downarrow}$), but no well-defined valley
polarization due to its strong localization. While also other specific
defect types have been proposed as possible origins of SPEs
\cite{Chakraborty2017,Zhang2017,Zheng2018}, we note that the presence
of any valley symmetry breaking defect seems sufficient, as long as it
results in a localized state near the defect site energetically close
enough to the bottom of the bulk conduction band to allow
hybridization due to strain [Fig.~\ref{fig:fig1}(a)].

The present one-particle description of a WSe$_2$ monolayer crystal that is locally
strained and decorated with a point defect is now the starting point
for inclusion of two-particle interactions. For the solution of the
Bethe-Salpeter equation (BSE) \cite{Salpeter1951} we employ our
one-particle wave functions to form a particle-hole basis
$\ket{c_{j},v_{i}} = \ket{c}^e \bigoplus \ket{v}^h$, with particle
state $\ket{c_j}$ and hole state $\ket{v_i}$, where the index ($i$,$j$)
refers to the valleys ($K,K'$) the states are associated with. Since
the excitonic states of interest are energetically well separated from the
conduction and valence band continua and spatially localized, we restrict ourselves to the
two-particle space spanned by,
\begin{subequations}\label{space}
\begin{equation}\label{space_a}
    \{ |c_{K} v_{K}\rangle, 
    |c_{K'} v_{K}\rangle, |c_{K} v_{K'}\rangle,|
    c_{K'} v_{K'}\rangle\} 
\end{equation}
and furthermore include defect excitons
    \begin{equation}\label{space_b}
  \{ |d_{\uparrow} v_{K}\rangle, |d_{\downarrow} v_{K}\rangle,
    |d_{\uparrow} v_{K'}\rangle, |d_{\downarrow} v_{K'}\rangle\}.
\end{equation}
\end{subequations}
 We solve the BSE-type equation
\begin{equation}
    \mathcal{H}^{\mathrm{tp}}|c_{i},v_{j}\rangle = 
    \mathcal{E} |c_{i},v_{j}\rangle,\quad
    \mathcal{H}^{tp} = (\mathrm{\epsilon}_{c_{i}} - 
                        \mathrm{\epsilon}_{v_{j}})
                        \mathrm{\delta}^{i}_{i'}
                        \mathrm{\delta}^{j}_{j'} +
    \Xi_{c_{i},v_{j}}^{c_{i'},v_{j'}}
    \label{Htp}
\end{equation}
where $\Xi_{c_{i},v_{j}}^{c_{i'},v_{j'}}
=W_{c_{i},v_{j}}^{c_{i'},v_{j'}} - V_{c_{i},v_{j}}^{c_{i'},v_{j'}} $
is the BSE interaction kernel, $W$ is the direct part and $V$ the
indirect contribution \cite{Reining2016} (for details, see SM).  The
direct part $W$ of the two-particle interaction shifts the states
downwards in energy by $\approx 100-500$ meV depending on the value
chosen for the dielectric constant (we use $\varepsilon/\varepsilon_0
= 10$ in the following \cite{Laturia2018}). Shifts of this order of
magnitude are consistent with experimentally observed excitonic
binding energies. While the indirect contribution $V$ is at least two
orders of magnitude smaller, it is key to understand the
fine-structure of SPE spectra. The direct term $W$ does not lift the
degeneracy since spin/valley locking allows only for non-vanishing
Hartree-like diagonal terms. In the absence of defects, spin/valley
locking prohibits also any non-vanishing off-diagonal contributions
for $V$ for strain-localized excitonic states $\{\ket{c,v}\}$ (In
contrast to bright A-excitons \cite{Glazov2014}).  Only in the
presence of defects with particle-hole states $\{\ket{d,v}\}$
off-diagonal contributions and, thus, fine-structure splittings of the
excitonic states (as observed in experiment) arise. The following
scenario for bright excitons emerges: diagonalizing the BSE
Hamiltonian [Eq.~(\ref{Htp})] in the subspace of Eq.~(\ref{space_b})
thereby neglecting the hybridization between the defect state and the
conduction band yields localized inter-valley defect excitonic (IDE)
states approximated by
\begin{equation}\label{eq:IDE}
  \ket{\mathrm{IDE}_\pm} \approx \frac{1}{\sqrt{2}}\left(\vphantom{\frac 12}\ket{d_{\uparrow(\downarrow)} v_{K(K')}}\pm\ket{d_{\downarrow(\uparrow)} v_{K'(K)}}\right),
\end{equation}
these IDE states appear in doublets ($\pm$) with an energy splitting of
$\rm{\Delta}_{0} \approx 0.8-2$ meV, well in the experimentally
observed range. Thus, the defect breaking the valley symmetry leads to
the formation of doublets [Eq.~(\ref{eq:IDE})] with an energy spacing
given by the exchange splitting. We note that inclusion of the
hybridization of defect states with the conduction band by
diagonalization of the BSE Hamiltonian in the full $8\times8$ space
[Eq.~(\ref{space})] can give rise to pairs of coupled doublets,
possibly accounting for recent observations \cite{He2016,Lu2019} (see
SM).

The IDE excitons [Eq.~(\ref{eq:IDE})] are efficiently populated by the
locally varying strain that shifts free ``bulk'' excitonic states
$\ket{c,v}$ in energy towards defect excitonic states $\ket{d,v}$
[Fig.~\ref{fig:fig1}] thereby effectively funneling population into
IDEs. Most importantly, the formation of defect excitons is accompanied
by a dramatic increase in optical transition strength (or reduction in
radiative lifetimes) when $\ket{c,v}$ and $\ket{d,v}$ approach each
other in energy [Fig.~\ref{fig:fig2} (d)].  While the transition
strength of ``bulk'' excitons $\ket{c,v}$, even in the presence of
strain, is of the order of $10^7\;\mathrm{s}^{-1}$ and thus too small
to serve as efficient photon emitter, the hybridization with
the defect state, which breaks the valley locking, increases the
transition strength by about two orders of magnitude to
$10^{9}\;\mathrm{s}^{-1}$.  The corresponding radiative lifetime,
which is of the order of nanoseconds, is in good agreement with
experiment.  These predictions are robust against variations of the
defect model or the strain pattern. In turn, spatially separating the
defect from the strained region decreases the transition rate as the
overlap between strain-localized excitons $\ket{c,v}$ and the
excitonic defect state $\ket{d,v}$ decreases.
\begin{figure}
    \centering
    \includegraphics[width=0.49\textwidth]{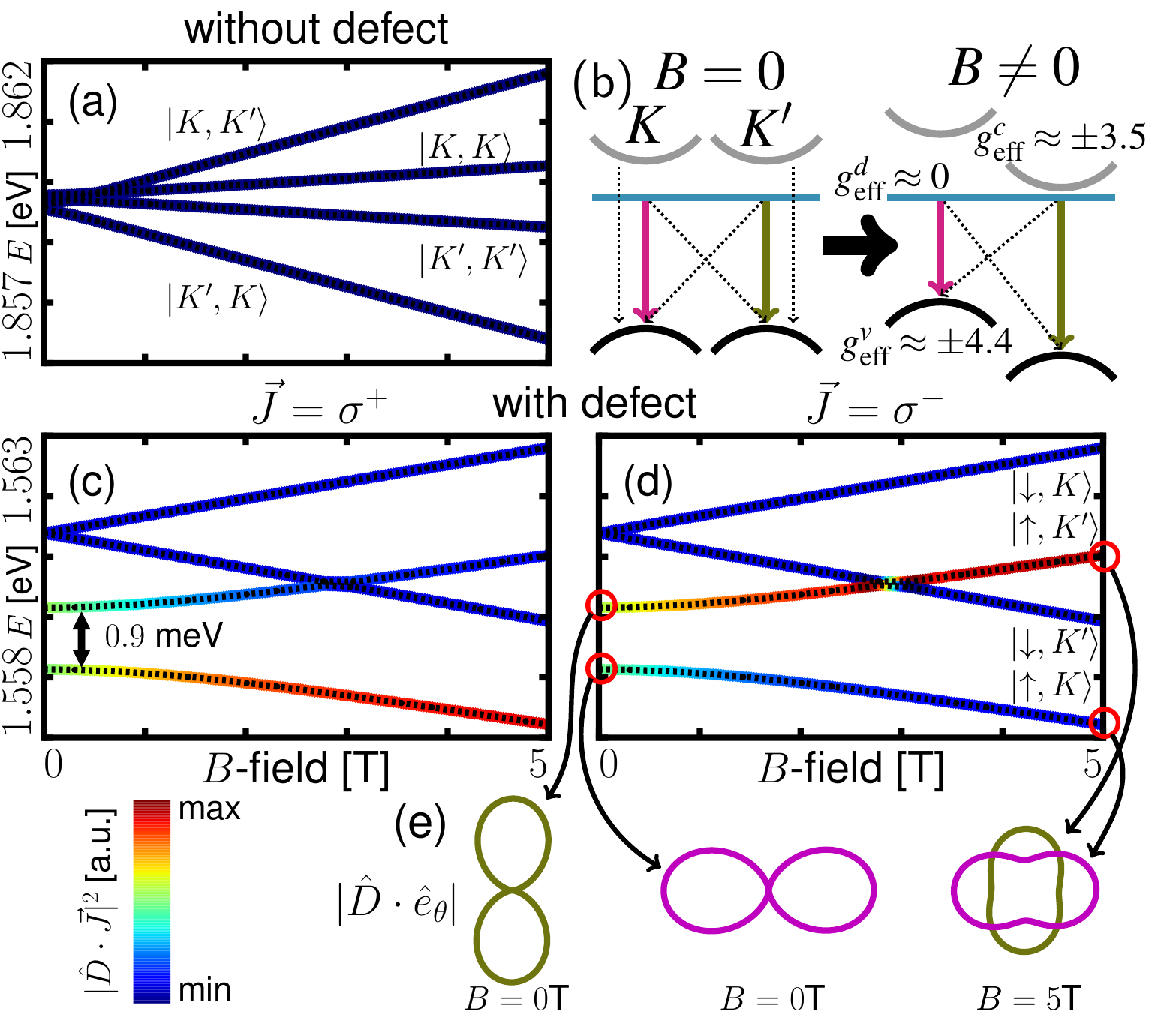}
    \caption{Magnetic field dependence of energy and polarization of
      localized excitons.  (a) without a defect.  (b) Schematic
      illustration of possible transitions and the calculated
      effective g-factor of single particle states.  All transitions
      are dark (dashed arrows) for the defect free case, while IDEs
      are optically active (colored arrows).  Simple Zeeman splitting
      of $\ket{c_{K(K')}}$ and $\ket{v_{K(K')}}$ in the absence of
      strain, together with the corresponding effective $g$-factors.
      (c)-(d) double Se vacancy (see SM for single vacancy). (c)
      right-handed [(d): left-handed] circularly polarized emission
      $\sigma^{+}$ [$\sigma^{-}$] as function of magnetic
      field. Color-scheme of the line marks the intensity of the
      transition by projection on the corresponding polarization vector
      $\vec{J} = {\sigma^+}$ [$\vec{J}={\sigma^-}$]. (e) Polar plot of
      the linear polarization of the emitted light as function of
      polarization angle.}
    \label{fig:fig3}
\end{figure}

The present model of localized IDE states as source of SPEs allows to
make detailed predictions for the response, both in energetic position
and polarization to magnetic and electric fields without resorting to
any adjustable parameter. 

With increasing magnetic field perpendicular to the crystal (Faraday
configuration) the zero-field exciton formed near a Se vacancy defect
undergoes a well known pronounced avoided crossing [Fig.~\ref{fig:fig3}(b)] with
splitting $\Delta(B) = \sqrt{\Delta_0^2 + (\mu_0
  g_{\mathrm{eff}}B)^2}$ and $\Delta_0$ the zero-field splitting of
the IDE doublet [Eq.~(\ref{eq:IDE})], in excellent agreement with
several measurements of the magnetic field evolution of SPE doublets
\cite{He2015,Srivastava2015,Chakraborty2015,Kumar2015,Clark2016,Branny2017,
  Chakraborty2017,Schwarz2016,He2016,Kumar2016,Berraquero2017,Luo2018}. 
The linearly polarized exciton at $B=0$ [right-handed ($\sigma^+$) and
  left-handed ($\sigma^-$) emission being equal] approaches circular
polarization with increasing magnetic field [Fig.~\ref{fig:fig3}(c,d)]
as for the free exciton \cite{Cao2012}.  Above $\approx 2$ T these
high-field excitons can again be associated with well defined valley
quantum numbers. In the high-field regime the magnetic response
becomes linear controlled by the orbital magnetic moment of the
valence (conduction) band states near the $K$($K'$) points $\mu \pm
4.4 \mu_0$ ($\pm 3.5 \mu_0$) for $\ket{v_{K(K')}}$ with opposite
signal for the two valleys, as they are connected by time reversal
symmetry (similar to \cite{Srivastava2015a}).  By contrast, a defect
strongly localizes on a few atomic sites and hardly contributes to the
shift with magnetic field. Therefore, the defect exciton with an
effective $g$-factor of $g_{\mathrm{eff}}=2 \cdot 4.4 = 8.8$ displays
a much smaller (larger) Zeeman shift than the bulk inter-valley
exciton $\ket{c_{K},v_{K'}}$ (intra-valley excitons
$\ket{c_{K},v_{K}}$) with an $g_{\mathrm{eff}}=2 \cdot (4.4 + 3.5) =
15.8$ ($g_{\mathrm{eff}}=2 \cdot (4.4 - 3.5) = 1.8$) [see
  Fig.~\ref{fig:fig3} (a,b)].  A g-factor of 15.8 was recently
reported for localized states in valley-aligned TMD heterobilayers \cite{Xu2019},
further underpinning our calculations.

We could not yet identify a systematic pattern that would connect the
polarization axis with the lattice orientation or the strain
gradient. However, our model allows for defining predictions for the
correlation between the polarization axes of SPEs residing in close
spatial proximity on the flake: the two lines from the doublet
(IDE$_\pm$) have polarization axes orthogonal to each other
[Fig.~3(e)] while excitons stemming from different inter-gap states of
one single defect feature linear polarization with the polarization
axis parallel to each other, as the lattice distortions (and therefore
the relative weights of the dipole matrix elements) are similar. 
  These
results suggest an explanation for the seemingly contradicting
measurements regarding either parallel or orthogonal relative linear
polarizations of spatially close SPE peaks. 
  
\begin{figure}
\centering
    \includegraphics[width=0.49\textwidth]{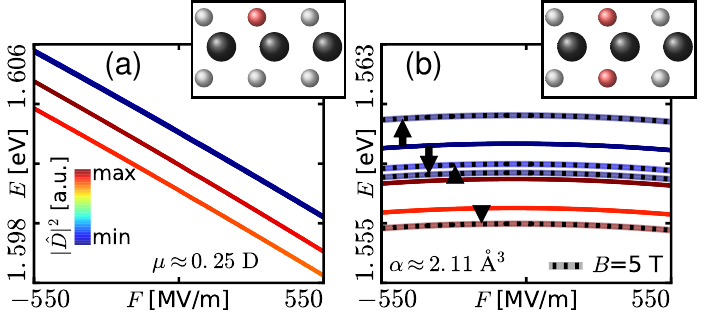}
    \caption{Stark shift of SPE. 
        (a) For a single Se vacancy (inversion symmetry breaking defect). 
        (inset) side view of the $\rm{WSe}_{2}$ layer with a vacancy (red).
        Color scheme as in Fig.~\ref{fig:fig3}
        (b) For a double Se vacancy (symmetry preserving) defect,
        dashed lines are calculated with an additional magnetic field 
        (5T).
        Black arrows indicate the evolution of each state with magnetic field.}
    \label{fig:fig4}
\end{figure}

Turning now to the dielectric response to an external electric field $F$
oriented perpendicular to the plane of the WSe$_2$ crystal, a wide array of
different experimental results have been reported. Parametrizing the
energy shift of the SPE as $E=E_0 - \mu_F F - \frac 12 \alpha F^2$
with $\mu_F$ the electric dipole moment and $\alpha$ the polarizability,
both linear and quadratic Stark shifts have been observed with $\mu_F$
ranging from $0.05$ to 10 Debye and $\alpha$ from 0.1 to 1000
$\mathrm{\AA}^3$ \cite{Chakraborty2017,Chakraborty2019,Schwarz2016}. 
The two prototypical point defects treated by our model, 
the single Se vacancy breaking the out-of plane inversion
symmetry and the double Se vacancy preserving this symmetry,
pinpoint the origin of such diverse results. 
With the out-of-plane inversion symmetry broken
by a single Se vacancy, we find a pronounced linear Stark shift with
$\mu_F = 0.25$ D [Fig.~\ref{fig:fig4} (a)]. For a double vacancy we observe only a quadratic
Stark effect with $\alpha=2.1 \; \mathrm{\AA}^3$ [Fig.~\ref{fig:fig4}]. 
Our results fit well to a
linear-response model which estimates the Stark shift based on the
density difference in the top and bottom Se layer.
For the present single Se vacancy the density asymmetry is about $1\%$, 
in principal allowing $\mu_F$ to be up to two orders of magnitude larger, 
well within the experimental range.\\

In summary, we have developed a microscopic model for bright single 
photon emitters in WSe$_2$ and have identified inter-valley defect 
excitons as likely candidates for strong photoemission. 
The interplay between strain and point defects allows to effectively 
funnel bulk excitons near the $K$($K'$) point into localized defect excitons. 
The broken lattice symmetry by the point defect breaks the spin-valley 
locking thereby opening the door to a large optical transition strength, 
a key prerequisite for bright photon emission. The broken valley symmetry 
also gives rise to an inter-valley mixture of the defect exciton, explaining 
the splitting in doublets at zero magnetic field. 
The predicted dielectric and paramagnetic response of the inter-valley 
localized defect excitons is consistent with a large number 
of experimental observations. The model is also capable of predicting 
the variation of the polarization of the SPE photons with 
applied magnetic field. Some intriguing questions, however, remain open. 
Among them are the statistics of energy and brightness fluctuations of the SPE, 
the conclusive identification of the dominant defect type, 
and the kinetics of the repopulation by the funnel. 
Addressing these questions is key to controlling single photon 
emission from WSe$_2$ for quantum optics and quantum information applications.
 
\acknowledgements

We acknowledge support by the TU-D doctoral program of TU Wien, as well as
from the Austrian Science Fund (FWF), project I-3827.

\clearpage
\newpage

\widetext
\begin{center}
\textbf{\large Supplemental Materials: Localized inter-valley defect excitons as single-photon emitters in WSe$_2$}
\end{center}
\setcounter{equation}{0}
\setcounter{figure}{0}
\setcounter{table}{0}
\setcounter{page}{1}
\makeatletter
\renewcommand{\theequation}{S\arabic{equation}}
\renewcommand{\thefigure}{S\arabic{figure}}
\renewcommand{\bibnumfmt}[1]{[S#1]}
\renewcommand{\citenumfont}[1]{S#1}

\section{DFT Calculations}
DFT calculations were performed using VASP \cite{Kresse1993,Kresse1994}.  We used
an orthorhombic unit cell containing 6 atoms per cell, a $k$ mesh of
19$\times$13$\times$1 $k$ points, 35 $\rm{\AA}$ vacuum in $z$ direction, PBE functional
and a non-collinear spin polarized basis.  All calculated
configurations were first calculated via ionic relaxation until
$\Delta F < 10^{-7}$.  Subsequently we include uniform strain in $x$
and $y$ direction, with 20 configurations in each direction ranging
from -2\% to 2\% strain.

The few highest valence bands and lowest conduction bands were then
transformed into an localized basis using
Wannier90 \cite{Marzari1997,Souza2001}.  As initial projections for the
Wannier orbitals we used the $4\rm{p}$ orbitals of Se and the
$5\rm{d}$ orbitals of W, each spin polarized.  While this choice of
initial projection results in a rather large tight-binding basis (6
sites per Se atom and 10 sites per W atom) it resembles the orbital
character of the system and therefore allows for a simple intuitive
picture.  The new localized basis still reaches DFT accuracy. \\ To
allow for a more direct comparison of involved energies, a scissor
operator was included in the transformation to obtain a direct band
gap of 2.0 eV in the unstrained case, close to the experimentally
observed electronic band gap \cite{He2014}. Including the scissor
operator required small modifications to the Wannier90 code.\\
\section{Non-uniform strain}

The DFT calculation yields hopping parameters $\gamma_{ij}$ between
all orbitals of the tight-binding basis of the WSe$_2$ lattice as a
function of uniform strain amplitudes. For inclusion of non-uniform
strain, each hopping parameter $\gamma_{{i},{j}}(\vec{r})$ between two orbitals
$i$ and $j$ separated by $\vec{r}= (x,y)$ is
approximated by a linear expansion of their spacing relative to the
relaxed configuration $\vec{r}_{0} = (x_{0},y_{0})$ (see [Fig.~S1 (a)]) 
\begin{equation}\label{eq:gammar}
    \gamma_{i,j}(\vec{r}) = \gamma_{i,j}^{0} + 
    (x-x_{0})\frac{\partial \gamma_{i,j}}{\partial x} + 
    (y-y_{0})\frac{\partial \gamma_{i,j}}{\partial y}.
\end{equation}
Since the overlap between neighboring orbitals decays exponentially, for 
small displacements ($|\vec{r}-\vec{r}_{0}| / |\vec{r}_{0}|\lesssim 3\%$) a linear  
approximation achieves a high accuracy.\\
This approach has several advantages: it does not require any additional fitting parameter, 
it preserves the original atomic orbital and spin configuration of the system thereby 
allowing for a straightforward physical interpretation and 
it sets no restrictions on the shape of the non-uniform strain pattern.
We are not aware of any other model for TMD's that can capture non-uniform strain configurations.
To test this description, we compare calculated bandstructures and 
hopping parameters obtained from an interpolation and from a direct DFT calculation 
and find almost perfect agreement (see [Fig.~S1 (b)-(e)]). 

\begin{figure}
\begin{center}
    \includegraphics[width=0.4\textwidth]{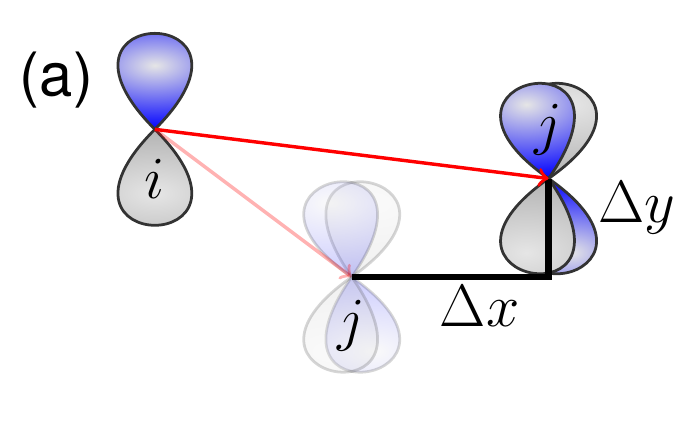}
    \includegraphics[width=1.\textwidth]{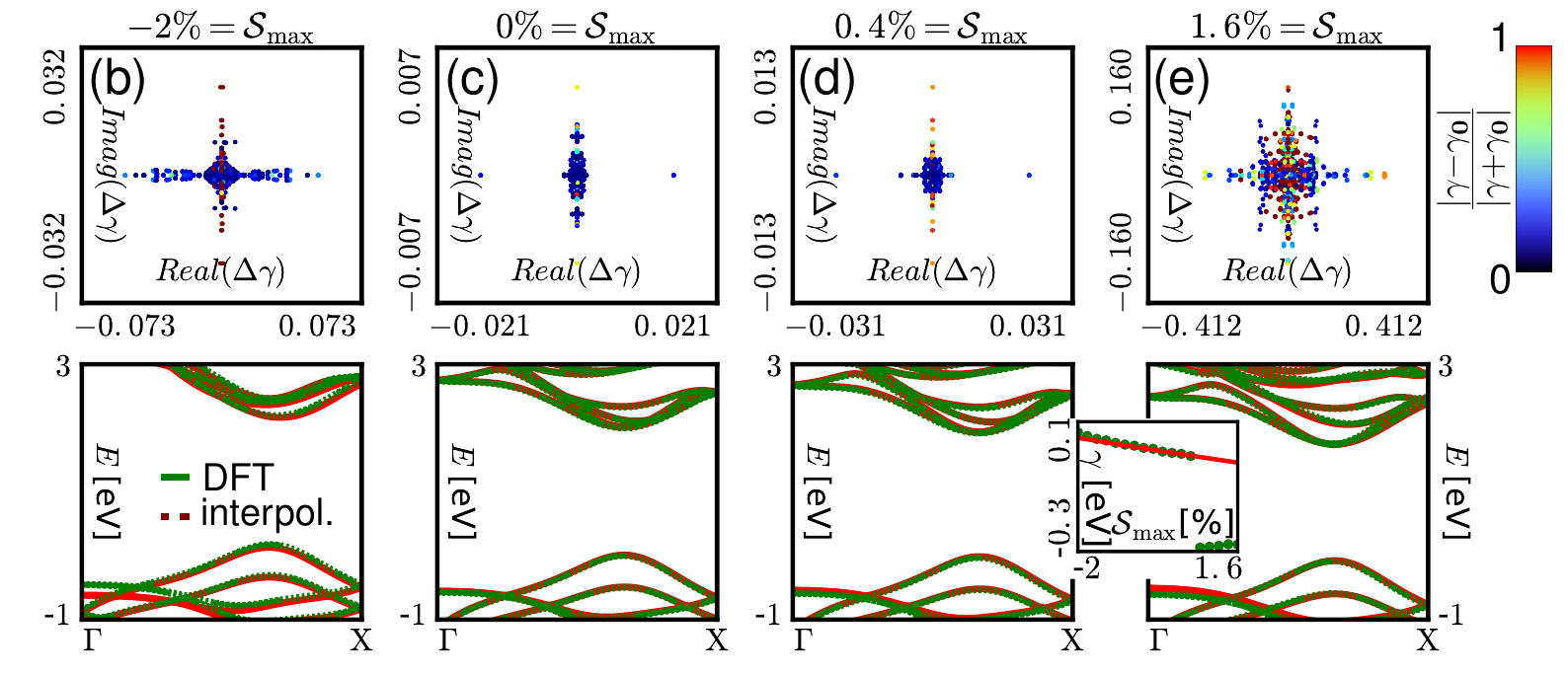}
\end{center}
\caption{Illustration of the interpolation of the interaction parameters 
    between orbital site $i$ and $j$. 
    (a) Schematic illustration of the interpolation.
    (b) first row: absolute and relative error of all interaction parameters 
    $\gamma_{i,j}$, between DFT calculated $\gamma$'s 
    and interpolated $\gamma$'s, for an orthorhombic unit cell strained in $x$ and $y$ direction.
    (b) second row: comparison of band structures for different strain values.
    (b) inset: example of one $\gamma$ parameter for different strain values. 
    Green dots indicate values from the Wannierization procedure and the 
    red line the interpolated values. 
    For large strain values ($\leq 1.4\%$) the Wannierization procedure 
    ends up in a slightly different minimum, 
    leading to large differences in the compared parameters.
    However this does not limit the accuracy of the description, 
    as the $\gamma$'s of the old Wannier minima can be linearly extrapolated, 
    still reproducing the band structure.}
\label{fig:fig_strain_explain}
\end{figure}

To simulate the localization by non-uniform strain, we use a large WSe$_2$ flake (38$\times$32 nm,
$\approx$ 200000 orbitals) with a Mexican-hat shaped strain pattern,
\begin{equation}\label{eq:svecr}
    \mathcal{S}(\vec{r}) = \mathcal{S}_{\rm{max}} \cdot 
    \bigg( 1 - \frac{|\vec{r}-\vec{r}_{0}|^2}{\rm{\sigma}^2}  \bigg) 
    \cdot \exp{ \bigg \{ -\bigg(\frac{|\vec{r}-\vec{r_{0}}|}{2\rm{\sigma}}\bigg) \bigg \} ^2}.
\end{equation}
with the strain maximum $\mathcal{S}_{\rm{max}}$ in the center of the flake
$\vec{r}_{0}$, and $\rm{\sigma}$ one third of the geometry width
[Fig.~S2].  The interaction parameters between
adjacent sites are now determined from our tight-binding model using
Eqs.~(\ref{eq:gammar}) and (\ref{eq:svecr}).  This strain pattern is
the simplest configuration that leaves the positions at the borders of
the geometry unchanged, allowing to attach open boundary conditions
included via self-energies of half infinite leads on all four edges of
the geometry \cite{Sanvito1999,Libisch2012,Papior2016} (see [Fig.~S2]).

\begin{figure}
\begin{center}
    \includegraphics[width=0.4\textwidth]{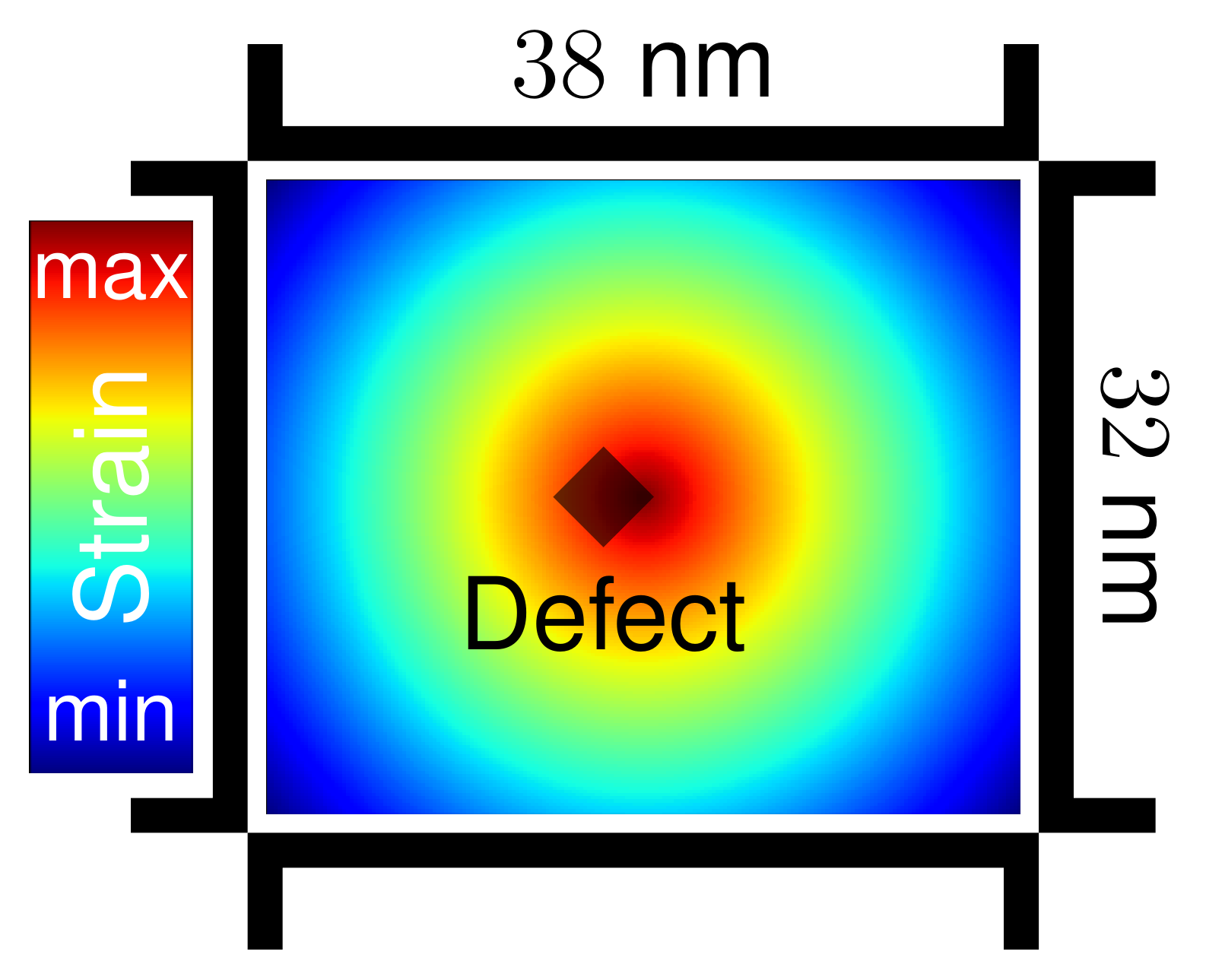}
\end{center}
    \caption{Schematic illustration of
      the model system containing a local strain maximum and a defect
      site (diamond).  Black lines indicate open boundary conditions calculated 
      via the self energies of half infinite leads \cite{Libisch2012} 
      and expanded via a Bloch-Ansatz \cite{Papior2016}. }
\label{fig:fig_geometry_explain}
\end{figure}
The single-particle eigenvalue problem can then be written as,
\begin{equation}
    \bigg(\mathcal{H}_{\mathrm{sp}} 
    + \mathrm{\Sigma}^{v(c)}_{\mathrm{L}}
    + \mathrm{\Sigma}^{v(c)}_{\mathrm{R}}
    + \mathrm{\Sigma}^{v(c)}_{\mathrm{T}}
    + \mathrm{\Sigma}^{v(c)}_{\mathrm{B}}
    \bigg)\Psi = \mathrm{\epsilon} \Psi,
\end{equation}
where $\Psi$ is the single particle wave function, 
$\mathcal{H}_{\mathrm{sp}}$ the single-particle Hamiltonian
and the $\Sigma$'s are the complex self-energies accounting for the 
openness of the flake in all for directions $\pm x$, $\pm y$.
These self-energies lead to a non-norm-conserving Hamiltonian (eigenenergies $\in \mathbb{C}$), 
the imaginary part of the eigenenergies introduces a measure for the 
coherent dissipation of the localized wave functions into the bulk. 
To efficiently calculate the self-energy for edges of this size we make use of 
the periodicity of the non-strained edge unit cells via a Bloch Ansatz \cite{Papior2016}.
The eigenstates $\Psi_{c(v)}$ around the valence band maximum and the conduction band minimum 
are calculated using a shift-and-invert Arnoldi-Lanczos scheme \cite{MUMPS,ARPACK}. 
The magnetic field is included via a Peierls phase \cite{Libisch2012}, 
defects are modeled by removing the orbitals at the defect site from the calculation geometry. 

Fig.~S3 (a) illustrates the evolution of the band structure as a
function of uniform strain.  While both band extrema move down in
energy, the conduction band minimum shifts much stronger with applied
strain than the valence state.  Employing a non-uniform strain pattern
leads to a localization of the conduction state with respect to the
maximum of the strain pattern $\vec{r}_{0}$ within the bandgap. The
valence band does not feature a similar localization on the single
particle level since the single particle states will be shifted
downwards in energy away from the bulk band gap [Fig.~S3 (a) valence
  band].  Due to a strong Coulomb interaction an excited electron-hole
pair will nevertheless localize at the electron site. Since we cannot
employ a self-consistent two-particle correction, we introduce a weak
localizing potential in the shape of the strain pattern with a maximum
energy $U_{\mathrm{max}}=80$ meV [see Eq.~(\ref{eq:svecr})], far smaller
than the excitonic binding energy. When introducing a localization
potential $\ket{v}$ is similar in shape to $\ket{c}$ [Fig.~S3 (right)
  (d)].

Qualitatively, the properties of the valence state wavefunction do not
change when including such a potential.  Moreover most of the
quantities discussed in this paper have numerical values independent
of the localization potential.  Only the calculated transition rate
$w$ and the magnitude of the zero field splitting
$\Delta_{0}$ change without localization potential, due to the
different electron-hole overlap.
However the relative transition rate between ''dark'' and ''bright'' states do not change, 
as well as the relative zero field splitting.
The valence state $\ket{v}$ is not
influenced by the presence of the defect (with and without
localization), since no defect state is close to the valence state
energy.

\begin{figure}
\centering
    \includegraphics[width=0.32\textwidth]{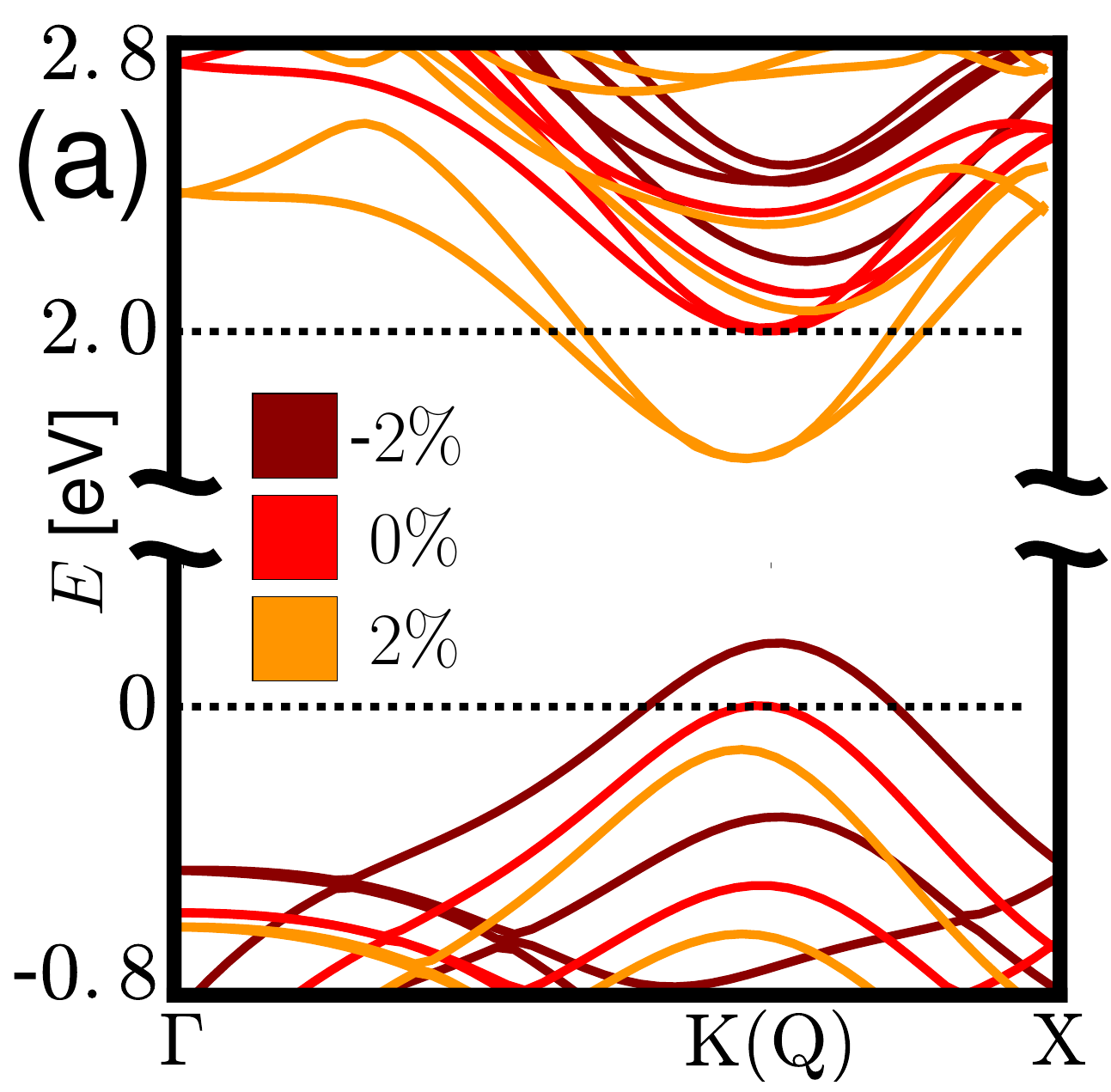}
    \includegraphics[width=0.4\textwidth]{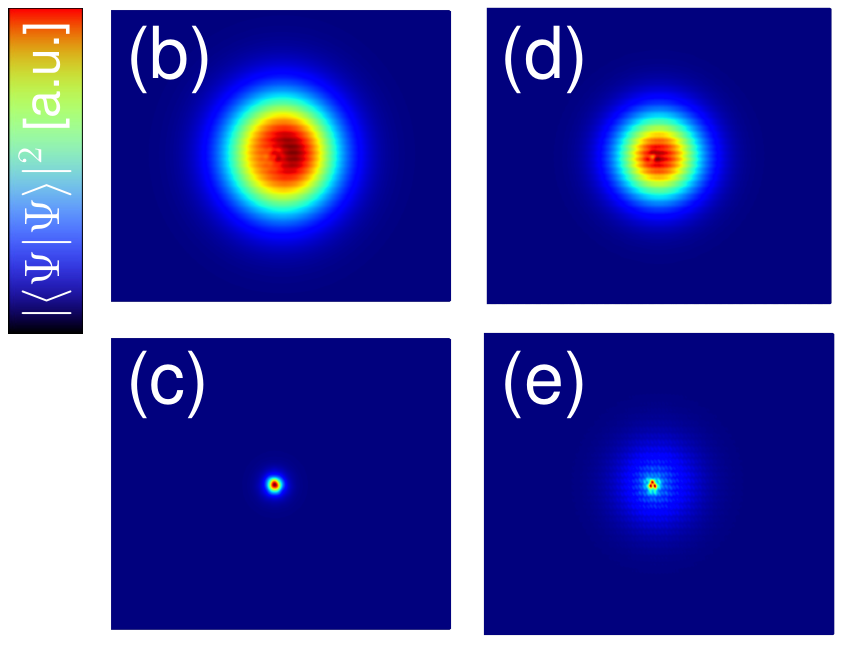}
    \label{fig:fig_strain_bs}
    \caption{
        (a) Calculated band structure of WSe$_{2}$ as a function of uniform strain. 
            Strain is applied uniformly in $x$- and $y$- direction. 
            For compressive strain we find WSe$_{2}$ to be an indirect semiconductor 
            ($\Gamma$-$\rm{Q}$ transition).
            For tensile strain we find a lowering of the conduction band 
            that is proportional to the applied strain, 
            with a slope of $\approx$ 140 meV/\% strain, 
            partly compensated by the valence band with $\approx$ 40 meV/\% strain.
            The conduction band splitting is strongly underestimated in DFT.
     (b) and (c) as in main text,
             (d) valence wave function when a localization potential is introduced,
             (e) defect wave function intermixed with the conduction state $\ket{c}$ 
             for a given strain configuration.}
\end{figure}

\section{Exciton transition rate}
The dominant decay channel for excitons are optical transitions with rate
\begin{equation}
    w = \tau^{-1} = \frac{\left(\mathrm{\omega}_{i\rightarrow f}^3
    | \vec{J} \cdot  \langle \Psi_{i}| \hat{\vec{r}}|\Psi_{f} \rangle \cdot q|^2\right)}
    {3\pi\mathrm{\varepsilon}_{0}\mathrm{c}^3\mathrm{\hbar}},
\end{equation}
with $\ket{\Psi_{f}}$ a valence wave function $\ket{v}$,
$\ket{\Psi_{i}}$ is $\ket{c}$ or $\ket{d}$, 
$\vec{J}$ the Jones vector and 
$\mathrm{\omega}_{i\rightarrow f} \hbar = \epsilon_{i} - \epsilon_{f} $.
Electron-phonon and electron scattering can be neglected as the
experiments are performed at cryostatic temperatures and the density
of accessible final states is low.
We note that we treat the two-particle interaction perturbatively and therefore 
neglect the change of transition rates due to two-particle interactions.
Furthermore, the considerations are restricted to $\Gamma$ only and we do 
not consider any additional decay channels.
Nevertheless we want to emphasize that relative transition 
rates between ``bright'' and ``dark'' excitons are expected to be captured by our model.

We calculate the position operator and subsequently the dipole transition rates via
\begin{equation}\label{eq:wannierdipole}
    \langle \phi_{i,\vec{R}} |\hat{\vec{r}}|\phi_{j,0} \rangle = 
    i \frac{\tilde{V}}{(s\pi)^{3}}\int d\vec{k} e^{i \vec{k}\cdot \vec{R}} 
    \langle u_{i,\vec{k}} | \nabla_{\vec{k}} |u_{j,\vec{k}} \rangle
\end{equation}
using the Wannier90 tool \cite{Marzari1997,Souza2001}, where
$\phi_{i,\vec{R}}$ is the Wannier basis function with index
$i$ in the unit cell translated by $\vec{R}$. 
We extend Eq.~(\ref{eq:wannierdipole}) to lattice site $\vec{r}$ via the
relation,
\begin{equation} \label{eq:wannierdipoleextend}
    \langle \phi_{i,\vec{R}'} | \hat{r} | \phi_{j,\vec{R}} \rangle =  
    \langle \phi_{i,\vec{R'}-\vec{R}} | \hat{r} | \phi_{j,0} \rangle + 
    \delta_{\vec{R},\vec{R}'}\delta_{i,j}\vec{R}.
\end{equation}
The approximation (\ref{eq:wannierdipoleextend}) does not account for
possible changes of the matrix element of $\vec{r}$ due to
inhomogeneous strain. The relative
shifts to neighboring atoms - i.e. the dominant contributions to the
position operator - are small and the strain influence as well as the defect relaxation 
are encoded in the wave functions. 
Therefore neglecting strain and  defect relaxation still provides
good estimates and correctly encodes information about valley and spin
transition rates. 

The optical oscillator strength can be interpreted in an ``excited'' 
exciton basis as the transition from
an exciton state to the exciton ground state,
\begin{equation}
    \langle 0,0 | \hat{D}_{c \rightarrow v}|v,c\rangle = 
    \langle \Psi_{i}| \hat{\vec{r}}|\Psi_{f} \rangle \cdot q.
\end{equation}

Under re-diagonalization, the dipole transitions of transformed exciton wave functions become
\begin{equation}
    \langle 0,0 | \hat{D} | \mathrm{IDE} \rangle = \sum_{n\in\{\uparrow,\downarrow\},  m\in\{K,K'\}}
    \alpha_{n,m}\langle 0,0 |\hat{D}_{n \rightarrow m} |d_{n},v_{m}\rangle.
\end{equation}
Therefore we obtain the dipole moment of the $\ket{\mathrm{IDE}}$ via a linear combination 
of the known excitonic transition rates of the defect excitons.

\section{Two-particle interaction}
The direct $W$ and indirect $V$ part of the interaction kernel 
$\Xi$ can be written in terms of the Coulomb operator as \cite{Reining2016}

\begin{equation}
     W_{c_{i},v_{j}}^{c_{i'},v_{j'}} =\frac 1 \varepsilon
    \langle c_{i},v_{j} | \hat{\mathrm{C}}_{i,j}^{i',j'} |c_{i'}, v_{j'} \rangle
\end{equation}
and
\begin{equation}
    V_{c_{i},v_{j}}^{c_{i'},v_{j'}} =
    \langle c_{i},v_{j} | \hat{\mathrm{C}}_{j',j}^{i',i}|c_{i'}, 
    v_{j'} \rangle.
\end{equation}
The Coulomb operator $\hat{\mathrm{C}}$ in real space representation reads,
\begin{equation}
    \hat{\mathrm{C}}_{m,n}^{l,k} = \sum_{\vec{r}\,,\vec{r'}\,'}
    \frac{\mathrm{\phi}^{\dagger}_{m}(\vec{r}\,)\mathrm{\phi}^{\dagger}_{n}(\vec{r}\,')
    \mathrm{\phi}_{l}(\vec{r}\,)\mathrm{\phi}_{k}(\vec{r}\,')}
    {|\vec{r}-\vec{r}\,'|}
\end{equation}
with $\mathrm{\phi}_{i}(\vec{r}) = \langle \vec{r} | v_{i}
(c_{i)}) \rangle$ and $s_{i}$ being the spin of state
$i$.  Note that the direct part is equivalent to the Hartree term
in a single-particle description while the indirect part corresponds, but is not equivalent, 
to exchange in a single particle description.
\newpage

\section{Magnetic field calculations for single Se vacancy}

Here we present the magnetic field dependent calculations of a single
Selen vacancy.  Since the single Se vacancy breaks the symmetry in z
direction, an additional avoided crossing is observed near two Tesla
where the lines cross for the double vacancy (see
[Fig.~S4 (a)-(c)]).\\ To underline the effects of
the direct contribution $W$ and the indirect contribution $V$ to the
magnetic field splitting we also show calculations including only the
direct contribution (see [Fig.~S4 (a), (d) and
(g)]).

\begin{figure}{h!}
    \centering
    \includegraphics[width=0.8\textwidth]{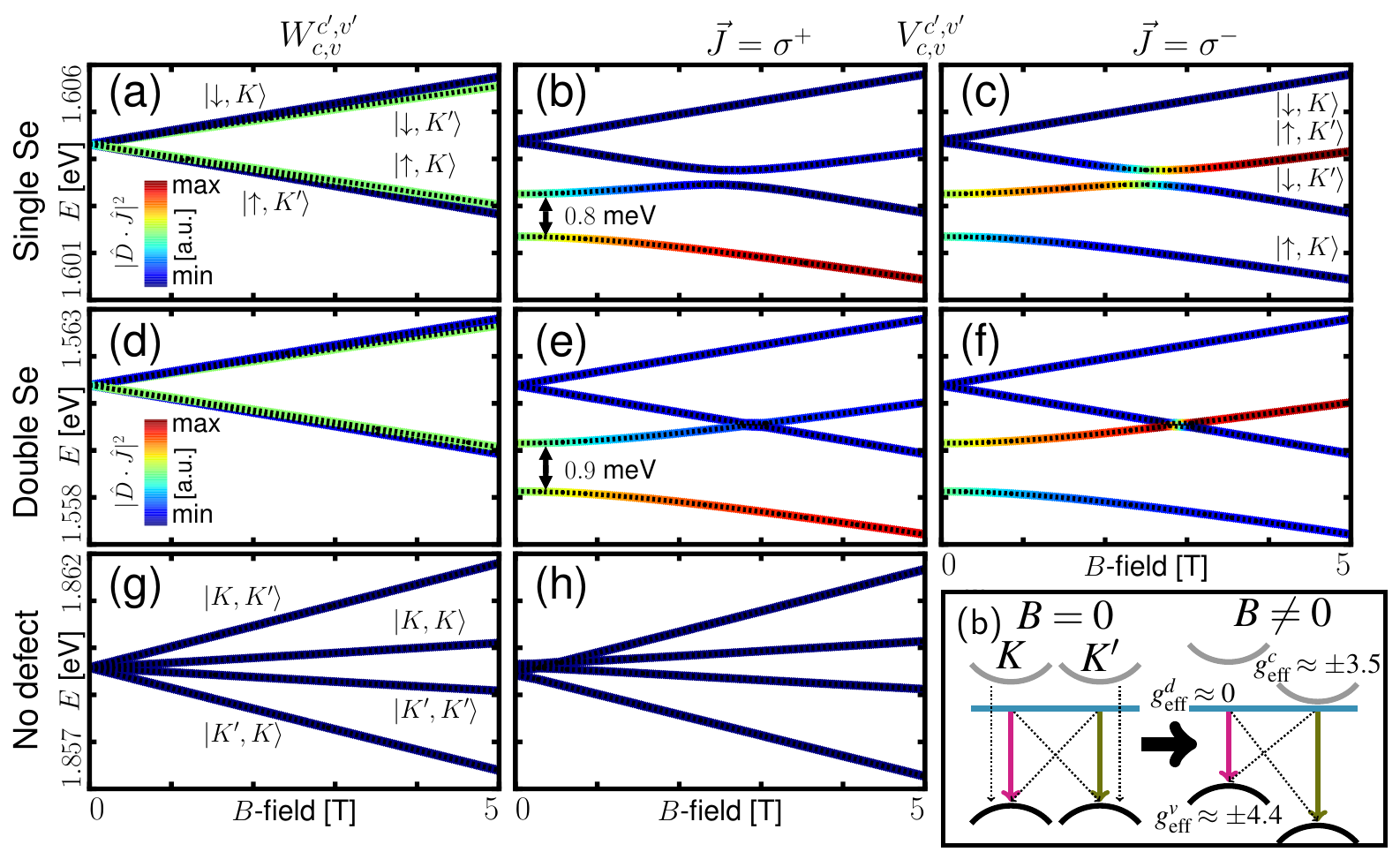}
    \caption{Plotted for comparison. 
             (a),(d) and (g) magnetic field splitting when only 
             the direct contribution $W$ is included.
             (a) for a single Se vacancy, (d) for a double Se vacancy and 
             (g) for a defect free system.
             (b)-(c) splitting when also including the indirect contribution $V$ 
             for a single Se vacancy, 
             (b) including right handed $\sigma^{+}$ and (c) left handed $\sigma^{-}$ light.
             When the symmetry in z-direction is broken by the single Se vacancy, 
             an additional avoided crossing is observed. 
             (e),(f) and (h) as in main text (see [Fig.~S3]).
             (i) pictorial illustration of the process at $B=0$ and at $B>0$.
             Cyan and brown arrows indicate allowed transitions, 
             dashed lines indicate ''dark'' transitions.
             Half-circles indicate $\ket{v}$ and $\ket{c}$ states, 
             the vertical line a defect state $\ket{d}$. The respective single particle effective 
             $g$-factors are also shown.
             }
    \label{fig:magnetic_all}
\end{figure}
\newpage

\section{Coupled doublets}

In the main text, we have given an approximate expression [Eq.~(3)]
for the localized inter-valley exciton $\ket{\mathrm{IDE}_\pm}$ in the limit
where the hybridization with the conduction band is neglected on the
BSE level. 
Considering now a stronger strain-induced mixture
between conduction band $\ket c$ and defect state $\ket d$, pairs of
hybridized states $\ket{h}_i$, $i=1,2$ emerge on the single-particle level
\begin{align}
    \ket{h}_{1,\mathcal{S}} &= \alpha({\mathcal{S}}) \ket{d_0} + \beta ({\mathcal{S}}) \ket{c_0}\quad
    \ket{h}_{2,\mathcal{S}} = \beta({\mathcal{S}}) \ket{d_0} - \alpha ({\mathcal{S}}) \ket{c_0}
\end{align}
with $\alpha^2 + \beta^2 = 1$, $\alpha (\mathcal{S}\rightarrow 0) =
1$, $\beta (\mathcal{S}\rightarrow 0) = 0$. Accordingly, the $\ket{\mathrm{IDE}}_\pm$
states evolve to the $\mathcal S \rightarrow 0$ limit of one of two hybridization states
\begin{equation}\label{eq:IDE}
  \ket{\mathrm{IDE}_\pm} \approx \frac{1}{\sqrt{2}}\left(\vphantom{\frac 12}\ket{d_{\uparrow(\downarrow)} v_{K(K')}}\pm\ket{d_{\downarrow(\uparrow)} v_{K'(K)}}\right).
\end{equation}
On the two-particle level, the hybridization is determined by the
interplay between strain and the Coulomb interaction. Consequently,
diagonalization of the full $8\times 8$ BSE Hamiltonian yields pairs
of localized inter-valley exciton doublets, $\ket{\mathrm{IDE}_\pm}^{(1)}$ and
$\ket{\mathrm{IDE}_\pm}^{(2)}$, weakly coupled and energetically several meV
apart.  The resulting fine-structure of these coupled doublets reflect
the mixture between defect excitonic $\ket{d,v}$ and bulk excitonic
$\ket{c,v}$ states. Accordingly, the zero-field splitting which is due
to the defect mode scales for the doublet $\ket{\mathrm{IDE}_\pm}^{(1)}$ as
$\Delta_1 \propto \alpha^2$ while for $\ket{\mathrm{IDE}_\pm}^{(2)}$ as
$\Delta_2 \propto \beta^2$ and will therefore decrease (increase) with
the degree of hybridization. Similarly, the transition rate will
depend on the relative weight of the defect admixture ($\propto
\alpha^2$ for $\ket{\mathrm{IDE}_\pm}^{(1)}$ and $\propto \beta^2$ for
$\ket{\mathrm{IDE}_\pm}^{(2)}$). The effective $g$-factors will be given by
$[g_{\mathrm{eff}}]^{(1)}=g_{\mathrm{eff}}^{d} \cdot \alpha^{2} +
g_{\mathrm{eff}}^{c} \cdot \beta^{2} $ and
$[g_{\mathrm{eff}}]^{(2)}=g_{\mathrm{eff}}^{d} \cdot \beta^{2} +
g_{\mathrm{eff}}^{c} \cdot \alpha^{2} $, where
$g_{\mathrm{eff}}^{d}=8.8\mu_{0}$ and
$g_{\mathrm{eff}}^{c}=15.8\mu_{0}$ (see main text). We believe this
dependence on the degree of hybridization explains the variations in
$g$-factors of single photon emitters reported in the
literature. Although
understanding this fine-structure needs further careful investigation,
these considerations show remarkable agreement with several
experimental observations (e.g. $\alpha=\beta$ \cite{He2016}, $\alpha
< \beta$ \cite{Lu2019}).

\section{Comparison with experiment}

Using our multi-scale tight binding model, we calculate a range of
properties of single-photon emitters.  The Table S1 presents our key
results and a comprehensive and concise comparison with experiment.
We include a wide range of experimental results available from the
literature. We find good agreement throughout. For observables such as
the effective $g$-factor for perpendicular magnetic fields or Stark
shifts, experiments yield a range of observed values that depend
on the precise level of defect hybridization, alignment with the
strain pattern, or symmetry breaking by the defect. Also for these
cases, we provide insight into the expected possible range of values
that is consistent with the spread found in the literature.

\begin{center}
    \begin{tabular}{l|c|c}
        Observable & Our model & Experiment \\ \hline \hline
        Energy [eV] &\bf 1.5-1.7& {\bf 1.55-1.72} \cite{He2015,Tonndorf2015,
                                        Chakraborty2015,Koperski2015,
                                        Kumar2015,Srivastava2015,Schwarz2016,Kumar2016,
                                        He2016,Clark2016,Kern2016,
                                        Berraquero2017,Chakraborty2017,Branny2017,
                                        Luo2018,Lu2019}\\
        Lifetime [ns] &\bf $\approx$ 1 & {\bf 0.5-8} \cite{He2015,Srivastava2015,Chakraborty2015,
                                              Koperski2015,Kumar2015,He2016,Kumar2016,Kern2016,
                                              Branny2017,Berraquero2017,Chakraborty2017}\\&&
                                              (0.1-0.5 Purcell enhanced) \cite{Luo2018}\\ \hline
        Strain & \bf pillar 1-3\%& {\bf pillar} \cite{Berraquero2017},
                                     { \bf rails} \cite{Kern2016}\\
               && measured \cite{Kumar2015,Kumar2016,Branny2017}\\ \hline
        g-factor [$\mu_{0}$]  &\bf  8.8 & 
                                      {\bf 7.16    } \cite{Kumar2015},
                                      {\bf 8.7} \cite{He2015}, 
                                      {\bf 7.7-10.9} \cite{Srivastava2015},\\&&
                                      {\bf 6.3     } \cite{Luo2018},
                                      {\bf 9-12    } \cite{Koperski2015},
                                      {\bf 9.8     } \cite{Chakraborty2015},
                                      {\bf 8       } \cite{Chakraborty2017}\\ &(9-14)&
                                      {\bf 9.4-10 (13)} \cite{Lu2019}\\
        $\Delta_{0}$ [meV] &\bf  0.8-2 & {\bf 0.7-0.9} \cite{He2015,Srivastava2015,
                                             Chakraborty2015,Kumar2015,
                                             Clark2016,Branny2017,
                                             Chakraborty2017},
                                     {\bf 1} \cite{Schwarz2016},\\&&
                                     {\bf 0.56} \cite{Lu2019},
                                     {\bf 0.2-0.4} \cite{He2016,Kumar2016},
                                     {\bf 0.2-0.73} \cite{Berraquero2017}\\ \hline
        Polarization&&\\
        \phantom{xxx}at $B=0$  & \bf linear   & {\bf linear} \cite{Tonndorf2015,He2015,
                                                                   He2016,Kumar2015,
                                                                   Lu2019,Schwarz2016,
                                            Kumar2016,Clark2016,Branny2017,Kern2016}\\
        \phantom{xxx}at $B>5T$ & \bf circular & {\bf circular} \cite{He2015,Srivastava2015,
                                                                     Lu2019,Koperski2015}\\ \hline
        Orientation&&\\
            \phantom{xxx}doublets  & \bf 90°      & {\bf $\approx$90°} \cite{He2015,Lu2019,
                                                  Schwarz2016,Clark2016,He2016}, 
                                     {\bf 40°-90°} \cite{Kumar2015}\\
            \phantom{xxx}two singlets & \bf $\approx$ 0° & {\bf 0°} \cite{Kumar2016,He2016}\\ \hline
        Stark shift&&\\ 
            \phantom{xxx}linear [D]&\bf 0.2 (0.1-10)& {\bf 0.058} \cite{Schwarz2016}, 
                                       {\bf 0.1-10} \cite{Chakraborty2017}\\
            \phantom{xxx}quadratic [$\rm{\AA}^{3}$] & \bf 2.1 & {\bf 1-1000} \cite{Chakraborty2017}
                                                                {\bf 3-98} \cite{Chakraborty2019} \\
            \hline

        Coupled doublets & \multicolumn{2}{c}{(see S VII.)} (\cite{Lu2019,He2016}) \\
    \end{tabular}
    \end{center}
    \label{table:table1}
    { \small TAB. S1: Comparison of our model to various experimental
      results. Values in brackets are estimated (theory)
      or their origin is unclear (experiment).}

\end{document}